\documentclass[11pt]{iopart}

\usepackage{graphicx,comment,siunitx,hyperref,ulem,float}
\usepackage{textcomp}
\usepackage[english]{babel}

\begin{document}

\title[Measuring spatial coherence of quantum and classical light]
{Measuring spatial coherence of quantum and classical light with an ultrastable monolithic interferometer}

\author{
Edoardo~Suerra$^{1,2}$,
Mirko~Siano$^{1,2}$,
Bruno~Paroli$^{1,2}$,
Samuele~Altilia$^{1,2}$,
Marco~A.~C.~Potenza$^{1,2}$,
Matteo~G.~A.~Paris$^{1,2}$,
and Simone~Cialdi$^{1,2}$}
\address{$^1$ Dipartimento di Fisica, Universit\`a di Milano, I-20133 Milan, Italy}
\address{$^2$ Istituto Nazionale di Fisica Nucleare, Sezione di Milano, I-20133 Milan, Italy}
\ead{edoardo.suerra@unimi.it}

\vspace{10pt}
\begin{indented}
\item[]July 2025
\end{indented}

\begin{abstract}
We describe a monolithic interferometer for spatial coherence measurements of both classical and quantum light sources. The design combines parametric down-conversion with a thermal source, using two identical calcite crystals to control beam alignment via birefringence. The monolithic structure ensures inherent stability. Spatial coherence is measured through temporal interferograms and spectral analysis, with both methods showing close agreement with theoretical predictions. The system is robust and performs reliably for both quantum and classical light. Its design enables automated, rapid coherence measurements across different source types.
\end{abstract}

%
%
%
%
%

\section{Introduction}
Spatial coherence is a fundamental property of both classical and non-classical light \cite{Loudon1980,Goodman1985,Allevi2014,Mandel1995}. 
It describes the ability of an electromagnetic field to maintain a fixed phase relation between different points across the beam profile \cite{Goodman1985}, and is traditionally quantified through the visibility of interference fringes in a Young’s double-slit experiment \cite{Young1804,Mandel1995}.
From a classical perspective, spatial coherence is of utmost importance in many research areas and applications, ranging from high-resolution optical microscopy \cite{Aguirre2015} to wavefront sensing \cite{Paroli_2018}, from coherent methods in the X-ray sciences \cite{Nugent01012010, Chapman_Nugent_2010} to free-space optical communications \cite{Paroli:21, PAROLI2022128808, terabit}. 
The coherence properties of the emitted light also carry useful information on the original radiation source, with applications such as measurements of stellar diameters \cite{stellar} and non-invasive particle beam diagnostics \cite{Siano2022}.
Spatial coherence becomes even more significant in the quantum regime, where it underpins remarkable phenomena such as quantum superposition and entanglement.  
In this context, the coherence properties of light are not merely classical features, but essential resources for quantum technologies, enabling groundbreaking advancements in quantum information, metrology \cite{Brida2010,Pittman1995}, and imaging \cite{Tan2008,Brida2010}. 
In particular, spatial optical correlations play a crucial role in quantum imaging \cite{Lugiato2002} and super-resolution \cite{Boto2000}, offering new possibilities for metrology, positioning, and high-precision measurements.
A striking example is their potential to surpass classical measurement limits; for instance, leveraging the spatial quantum correlations of spontaneous parametric down-conversion (SPDC) emission to detect weak objects could drive significant practical progress \cite{Brida2010}.

For these reasons, the characterization of spatial coherence — both in quantum and classical regimes — is of paramount importance, and robust and straightforward methods for measuring the spatial coherence of a source are highly desirable. 
From the quantum perspective, the spatial properties of light — particularly those of twin photons — were investigated as early as the first experimental observation of parametric down-conversion by Burnham and Weinberg \cite{Burnham1970}, who noted that intensity correlations were stronger for specific combinations of detection angles. 
Subsequent experiments, such as those exploring double-slit interference using twin photons and coincidence counting \cite{Ribeiro1994,Strekalov1995}, further elucidated these spatial correlations, with later studies providing more detailed analyses \cite{Fonseca1999,Walborn2010}. 
Additional methodologies, including modified Michelson interferometers \cite{Cutipa2020,Kwon2009}, have also been explored.
Numerous approaches have been proposed also for determining the spatial coherence of classical light, including techniques based on Young's interferometer \cite{Partanen2019,Saastamoinen2018}, wavefront folding and shearing interferometers \cite{He1988,Wessely1970,Breckinridge1972,Arimoto1997}, random speckle patterns \cite{Alaimo2009,PhysRevA.92.033842,PhysRevLett.108.024801,Siano2021}, gratings \cite{Pfeiffer2005}, plasmonic devices \cite{Morrill2016}, reversed-wavefront interferometers \cite{Santarsiero2006}, digital micromirror devices \cite{Partanen2014}, among others \cite{Turunen2022}. 
Typically, one of the main challenges in experimentally detecting and investigating quantum light lies in the intrinsically low power of quantum states, which often renders them nearly undetectable.
As a result, studying their spatial properties, for instance, by translating duplicate copies of the same beam, becomes experimentally impractical.
This necessitates the use of intrinsically aligned and highly stable detection systems.

In this work, we introduce an interferometric technique based on a monolithic interferometer to measure the spatial coherence of both quantum and classical light sources.
The proposed device, which incorporates two calcite crystals, features an inherently ultrastable design that enables precise control over both spatial and temporal beam overlap.
Its monolithic architecture ensures intrinsic alignment throughout the measurement process.
We demonstrate the effectiveness of this approach by characterizing the spatial coherence of two distinct light sources: a twin-photon source generated via spontaneous parametric down-conversion (SPDC) and a classical thermal source with a longitudinal coherence length on the order of \SI{1}{\micro\meter}.
The measured transverse coherence lengths are in excellent agreement with theoretical predictions, highlighting the robustness and accuracy of the method.

The paper is organized as follows:  
Section~\ref{sec:exp} describes the experimental setup in detail;  
Section~\ref{sec:theory} outlines the theoretical framework necessary to describe spatial coherence;  
Section~\ref{sec:res} presents the experimental results;  
finally, Section~\ref{sec:conc} concludes the paper with a discussion of the main findings.

\section{Monolithic interferometer and experimental setup}\label{sec:exp}
\begin{figure}
    \centering
    \includegraphics[width=82mm]{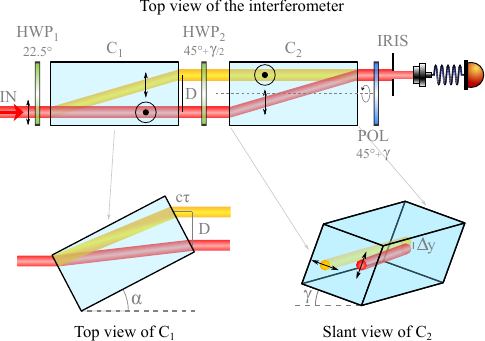}
    \caption{
    Schematic of the monolithic interferometer (top) and detailed views of the calcite crystals (bottom). IN: input beam; HWP$_i$: half-wave plates; C$_i$: calcite crystals; POL: polarizer; IRIS: iris diaphragm. Rotation of C$_1$ by $\alpha$ introduces a time delay $\tau$, while rotation of C$_2$ by $\gamma$ produces a vertical displacement $\Delta y$.}
    \label{fig:detector}
\end{figure}

The monolithic interferometer used in this study is illustrated schematically in Fig.~\ref{fig:detector}. 
It comprises two identical \SI{40}{\milli\meter}-long calcite crystals and follows the design detailed in Ref.~\cite{Cialdi2024}, itself based on the implementation of Ref.~\cite{Siano2024}.
In this work, we introduce for the first time the possibility of overlapping distinct transverse portions of the beam to probe its spatial coherence using this interferometer.
Initially, a beam — either quantum or classical (denoted IN) — is prepared in a horizontal polarization state. 
A half-wave plate (HWP$_1$) then rotates this polarization to \SI{45}{\degree}. 
The first calcite crystal (C$_1$) separates the vertical (ordinary) and horizontal (extraordinary) components: the extraordinary ray undergoes a spatial walk-off of $D = \SI{4.18}{\milli\meter}$ upon exiting C$_1$. 
A second half-wave plate (HWP$_2$) subsequently swaps the polarization components before the beams enter the second calcite crystal (C$_2$). 
After C$_2$, the two paths recombine and are projected onto a \SI{45}{\degree} polarizer (POL), producing an interference pattern. 
An iris (IRIS) restricts the detection to the desired spatial region.
The temporal delay $\tau$ between the two arms is controlled by rotating C$_1$ by a small angle $\alpha$ around its vertical axis (see top view in Fig.~\ref{fig:detector}). 
This rotation introduces an additional free-space path $\Delta z = c\,\tau = D\tan{\alpha} \approx D\,\alpha$ for the extraordinary beam, while any alteration of the internal crystal path due to Snell's law is negligible. 
Similarly, rotating C$_2$ by a small angle $\gamma$ around its optical axis induces a vertical shear $\Delta y = D\sin\gamma \approx D\gamma$, with a negligible horizontal shift $\Delta x = D(1-\cos\gamma) \approx 0$. 
To maintain optimal interference, HWP$_2$ and POL must be rotated consistently with the orientation of C$_2$, specifically by $\gamma/2$ and $\gamma$, respectively. 
In this way, transverse spatial coherence can be measured observing the visibility of the inteference pattern at the output of the interferometer as a function of the transverse shift.
Also, temporal coherence can be measured observing the visibility of the interference pattern at the output of the interferometer as a function of the temporal delay.

\begin{figure}
    \centering
    \includegraphics[width=82mm]{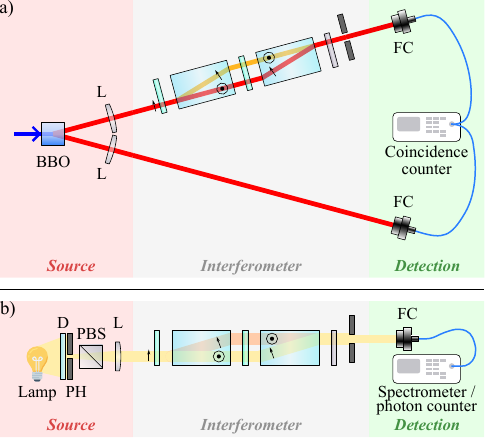}
    \caption{
    Layout of the coherence measurement apparatus using (a) quantum radiation generated by type-I spontaneous parametric down-conversion (SPDC) in a \SI{3.00}{\milli\meter}-long beta-Barium Borate (BBO) crystal, and (b) a classical thermal source. Acronyms: BBO, beta-Barium Borate; L, lens; FC, single-mode fiber coupler; D, diffuser; PH, pinhole; PBS, polarizing beam-splitter.}
    \label{fig:setup}
\end{figure}

Fig.~\ref{fig:setup} shows the overall experimental setup for measuring spatial coherence where we implemented our monolothic interferometer, with either quantum or classical light sources.
For the quantum configuration (Fig.~\ref{fig:setup}a), photon pairs are produced via type-I SPDC in a \SI{3.00}{\milli\meter}-long BBO crystal, pumped by a continuous-wave laser at \SI{405}{\nano\meter}. 
The pump beam is spatially filtered to ensure a Gaussian mode within the BBO, yielding signal and idler photons at \SI{810}{\nano\meter}. 
We tested two pump waists in the crystal of \SI{560}{\micro\meter} and \SI{840}{\micro\meter}. 
The idler photon is collimated by a \SI{500}{\milli\meter}-focal-length lens and detected by a single-photon counter, while the signal photon follows an identical collimation before entering the interferometer. 
An electronic timing circuit records coincidence events between signal and idler detectors.

In the classical arrangement (Fig.~\ref{fig:setup}b), a halogen lamp provides broadband light which is diffused by a ground-glass plate, spatially filtered through a \SI{1}{\milli\meter}-diameter iris, and horizontally polarized by a polarizing beam splitter. 
The resulting beam, centered at $\lambda_0=\SI{680}{\nano\meter}$ with a spectral bandwidth $\Delta\lambda=\SI{150}{\nano\meter}$ (FWHM), has an estimated longitudinal coherence length $\lambda_0^2/\Delta\lambda\approx\SI{3}{\micro\meter}$. 
After collimation by a \SI{500}{\milli\meter}-focal-length lens, the beam enters our monolithic interferometer. 
The spatially and temporally sheared replicas are either analyzed spectrally using a spectrometer — enabling direct determination of spatial coherence as detailed in Sect.~\ref{sec:theory} — or detected with a photon counter to extract both spatial and temporal coherence from interferograms. 
A computer controls two stepper motors used for C$_1$ and C$_2$ rotation, and acquires data automatically.

\section{Theoretical model}\label{sec:theory}
This section presents a theoretical analysis of the spatial coherence properties of both the quantum and classical sources used in our experiment. We begin by examining the spatial correlations of light generated through SPDC, specifically focusing on type-I SPDC, which is the process employed in our setup.

At the output of the nonlinear crystal, the two-photon quantum state can be written, up to multiplicative constants, in the well-known form:
\begin{equation}
    \left| \psi \right> =
    \int \mathrm{d}\omega \, 
    \mathrm{d}^2 \vec{k}_s \, \mathrm{d}^2 \vec{k}_i \,
    \tilde{A}_p\!\left(k_x,k_y\right) \, \mathrm{sinc}{\left(\frac{L}{2} \Delta k_z\right)} 
    \left|\vec{k}_s,\omega\right>_s \left|\vec{k}_i,-\omega\right>_i \, ,
\end{equation}
where a monochromatic pump field is assumed.
Here, $p$, $s$, and $i$ refer to the pump, signal, and idler beams, respectively. 
The transverse components of the wavevectors are denoted as $\vec{k}_j = (k_{x,j}, k_{y,j})$, and $\omega$ represents the angular frequency detuning of the down-converted photons relative to the central angular frequency $\omega_0 = \omega_p / 2$.
Moreover, $\tilde{A}_p$ is the Fourier transform of the pump field amplitude, and $\Delta k_z = k_{z,p} - k_{z,s} - k_{z,i}$ is the longitudinal phase mismatch.
The $\mathrm{sinc}$ term originates from the phase-matching function integrated over the crystal length $L$. In our configuration, this function is sufficiently broad that it can be approximated as unity.
Assuming a Gaussian pump beam with beam waist $w_p$, its angular spectrum is given by:
\begin{equation}
    \tilde{A}_p\!\left(k_x,k_y\right) =
    e^{-\frac{w_p^2 \left|\Delta\vec{k}_{\perp}\right|^2}{4}} = 
    e^{-\frac{w_p^2 \Delta\vec{k}_{\perp,x}^2}{4}}
    e^{-\frac{w_p^2 \Delta\vec{k}_{\perp,y}^2}{4}} 
    =
    \tilde{A}_{p,x} \tilde{A}_{p,y} \, ,
\end{equation}
where $\Delta\vec{k}_{\perp} = \vec{k}_s + \vec{k}_i$ is the total transverse momentum mismatch.
Substituting this expression into the quantum state, we obtain:
\begin{equation}
    \left| \psi \right> =
    \int \mathrm{d}\omega \, 
    \mathrm{d}^2 \vec{k}_s \, \mathrm{d}^2 \vec{k}_i \,
    \tilde{A}_{p,x} \tilde{A}_{p,y} 
    \left|\vec{k}_s,\omega\right>_s \left|\vec{k}_i,-\omega\right>_i \, .
\end{equation}

We now consider the effect of two identical lenses, each of focal length $f$, used to collimate the signal and idler beams. Each lens maps the transverse momentum $\vec{k}_\perp$ to a transverse spatial coordinate $\vec{r}$ via the relation $\Delta\vec{k}_\perp = \frac{2\pi}{\lambda f} \Delta\vec{r}$. After this transformation, the state becomes:
\begin{equation}
    \left| \psi \right> =
    \int \mathrm{d}\omega \,
    \mathrm{d}x \, \mathrm{d}y \,
    \tilde{A}_{p,x} \tilde{A}_{p,y} 
    \left|x_s,y_s,\omega\right>_s \left|x_i,y_i,-\omega\right>_i \, .
\end{equation}

As discussed in Section \ref{sec:exp}, our interferometric setup superposes two copies of the quantum state along the vertical ($y$) direction. We can therefore restrict our analysis to this coordinate. At the output of the interferometer, the state undergoes both a vertical displacement $\Delta y$ and a temporal delay $\tau$, followed by projection through a polarizer that allows the two components to interfere. The resulting state at the detector is given by:
\begin{equation}
    \left| \psi \right> =
    \frac{1}{2} \int \mathrm{d}\omega \,
    \mathrm{d}y \,
    \tilde{G}
    \left|y_s,\omega\right>_s \left|y_i,-\omega\right>_i \, ,
\end{equation}
where 
\begin{equation}
    \tilde{G} = \tilde{A}_{p,y}\!\left(y_s-y_i\right) + \tilde{A}_{p,y}\!\left(y_s-y_i - \Delta y \right) e^{i\,\tau\left(\omega_0+\omega\right)} \, .
\end{equation}

The coincidence detection probability is obtained by projecting this state onto position and frequency eigenstates:
\begin{equation}\label{eq:pdcprob}
    P(\tau,\Delta y) = 
    \int \mathrm{d}\omega'\,\mathrm{d}y'_s\,\mathrm{d}y'_i\,
    \left|
        {}_{s}\langle y'_s,\omega'| 
        {}_{i}\langle y'_i,-\omega'| 
        \psi \rangle 
    \right|^2 
    = \frac{1}{2} + \frac{1}{2} 
    g(\Delta y)
    \cos\!\left(\omega_0 \tau\right) \, .
\end{equation}
Here, the function 
\begin{equation}\label{eq:gquantum}
    g(\Delta y) = \int \mathrm{d}y_s\,\mathrm{d}y_i\,\tilde{A}_p\!(y_s-y_i)\tilde{A}_p\!(y_s-y_i-\Delta y)
\end{equation}
describes the spatial correlations between the photons. 
The $\cos(\omega_0 \tau)$ term gives rise to interference fringes, and the visibility of these fringes is determined by $g(\Delta y)$.
Therefore, the measurement of interference visibility — specifically the dependence on the vertical shift $\Delta y$—provides direct information about the spatial coherence and correlations in the photon pairs.

From the classical point of view, let us consider a thermal source of diameter $D$, as schematized in Fig.~\ref{fig:setup}b.
The thermal light is linearly polarized and collimated with a lens of focal length $f$, then it is directed into our interferometer.
As for the quantum counterpart, two copies of the beam are overlapped with a temporal shift $\tau$ and a spatial (vertical) translation $\Delta y$.
In the end, the radiation can be coupled either to a spectrometer or a photon counter.
Considering the spectrometer, the resulting spectrum is given by
\begin{equation}
    S(\lambda) = 
    S_0(\lambda)
    \left[
    1 + \mu(\Delta y) \cos\!\left(\frac{2\pi c \tau}{\lambda}\right)
    \right] \,,
\end{equation}
where interference fringes due to the temporal shift $\tau$ are modulated by a visibility that depends on the spatial shift $\Delta y$, which we can identify as the spatial coherence of the radiation. 
Thus, by measuring the fringes visibility in the spectrum, one can retrieve the coherence function $\mu$ of the source.
For the sake of simplicity, let us consider a circular source of radius $r$, whose coherence function at distance $f$ is given by \cite{Goodman1985}
\begin{equation}\label{eq:mucirc}
    \mu_\mathrm{circ}(\Delta y,\lambda)=
    2\left|\frac{
    J_1\left(\frac{2\pi r}{f \lambda} \Delta y\right)
    }{
    \frac{2\pi r}{f \lambda} \Delta y
    }
    \right| \, ,
\end{equation}
being $J_1\left(z\right)$ the Bessel function of the first kind.
This is the coherence function of our circular source after collimation \cite{Goodman1985}.
By measuring fringes visibility in the spectrum, one can retrieve information about source shape.
Since $\mu$, and thus visibility, depends on the wavelength $\lambda$, a more accurate analysis can be done by considering the reduced coordinate $\Delta \tilde{y} = \Delta y / \lambda$.
This simply gives a spectrum
\begin{equation}\label{eq:specred}
    \tilde{S}(\Delta \tilde{y}) = 
    1 + \mu(\Delta \tilde{y}) 
    \cos\!\left(\frac{2\pi c \tau}{\Delta y} \Delta\tilde{y}\right) \,,
\end{equation}
and, in case of a circular source, a coherence function
\begin{equation}\label{eq:mucirc}
    \mu_\mathrm{circ}(\Delta \tilde{y})=
    2\left|\frac{
    J_1\left(\frac{2\pi r}{f} \Delta \tilde{y}\right)
    }{
    \frac{2\pi r}{f} \Delta \tilde{y}
    }
    \right| \, ,
\end{equation}
from which one can retrieve information about source geometry (i.e., radius).
If we consider the photon counter instead of the spectrometer, we can acquire interferograms by varying $\tau$.
The normalized intensity pattern of the interferogram can be written as \cite{Goodman1985}
\begin{equation}\label{eq:cohetot}
    I\left(\tau,\Delta y\right) =
    1 + \mu_T\left(\tau\right) \mu_S\left(\Delta y\right)
    \cos\!\left(\frac{2\pi c}{\lambda_0} \tau + \phi\left(\Delta y\right)\right) \, ,
\end{equation}
where $\mu_T\left(\tau\right)$ and $\mu_S\left(\Delta y\right)$ are the temporal and spatial coherence functions, respectively.

It is important to highlight the effect of beam divergence on spatial coherence measurements, emphasizing the crucial role of beam collimation in the implementation of this method.
Let us consider a spherical wavefront with a radius of curvature $R$ at the entrance of the first crystal, C$_1$.  
Under the paraxial approximation, the two-crystal interferometer produces straight interference fringes at the detection plane with a periodicity given by $\Lambda = \frac{\lambda}{\Delta y} \left(R+d\right)$, where $d$ is the optical distance between the entrance of C$_1$ and the detector.  
Defining $\Phi$ as the diameter of the collection aperture of the detector, the impact of the interference fringes on spatial coherence measurements becomes negligible when $\Lambda > \Phi$.  
Conversely, if this condition is not met, the visibility of the interferograms or spectra will be affected. 
As an example, for typical experimental parameters such as $\Delta y_\mathrm{max} = \SI{1}{\milli\meter}$, $\Phi_\mathrm{max} = \SI{2}{\milli\meter}$, $\lambda = \SI{700}{\nano\meter}$, and $d = \SI{100}{\milli\meter}$, the constraint $R > \SI{2.7}{\meter}$ must be satisfied.  
This requirement must be fulfilled by introducing a collimation lens to ensure a well-collimated source, as implemented in our setup.

\section{Results}\label{sec:res}
\begin{figure}
    \centering
    \includegraphics[width=82mm]{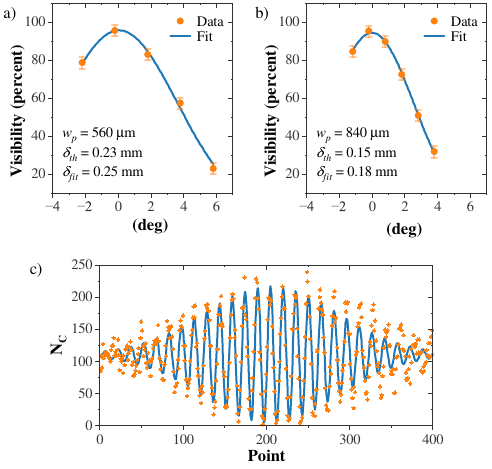}
    \caption{Maximum visibility of the interferograms as a function of spatial displacement $\Delta y$ for the quantum source from PDC, and for two different pump dimensions of \SI{560}{\micro\meter} a), and \SI{840}{\micro\meter} b).
    In c) we show an example of an interferogram on the coincidence photon countings $N_C$, where orange points represents experimental data, while the blue line is their fit.}
    \label{fig:resPDC}
\end{figure}

Here  we present the results obtained for the quantum  and the classical source.
Figs.~\ref{fig:resPDC}a and \ref{fig:resPDC}b reports the measurements of the fringe visibility in the coincidence photon counting as a function of the vertical separation of the states in the interferometer.
Orange points are the experimental data, while the blue line is a fit with the theoretical function given by $g(\Delta y)$ of Eq.~\ref{eq:gquantum}.
In Fig.~\ref{fig:resPDC}a the pump size is \SI{560}{\micro\meter} (beam radius $w$), giving a spatial coherence length of \SI{250}{\micro\meter}.
The theoretical value of \SI{230}{\micro\meter} is calculated simply by $\delta = \frac{\lambda_0 f}{\pi w}$, and it is in good agreement with the value retrieved by the fit.
In Fig.~\ref{fig:resPDC}b the pump size is \SI{840}{\micro\meter}, corresponding to a spatial coherence length of \SI{180}{\micro\meter}, again in good agreement with the theoretical value of \SI{150}{\micro\meter}.
Note that in these measurements dark counts are negligible.
The theoretical value is probably underestimated due to uncertainties of lens focal length and beam size.
These results demonstrate that this method can be successfully implemented for measurements of spatial correlations of a quantum source.
Fig.~\ref{fig:resPDC}c shows an example of an interferogram of the coincidence photon countings, with a vertical shift of \SI{40}{\micro\meter}.
Here, orange dots are experimental data, while the blue line is a fit with Eq.~\ref{eq:pdcprob}.

\begin{figure}
    \centering
    \includegraphics[width=82mm]{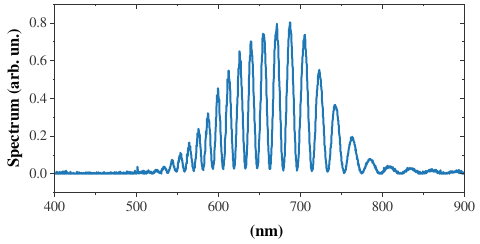}
    \caption{Example of measurement of the spectrum of classical thermal source, with a temporal delay of $\tau = \SI{93}{\femto\second}$.
    }
    \label{fig:specclas}
\end{figure}

\begin{figure}
    \centering
    \includegraphics[width=82mm]{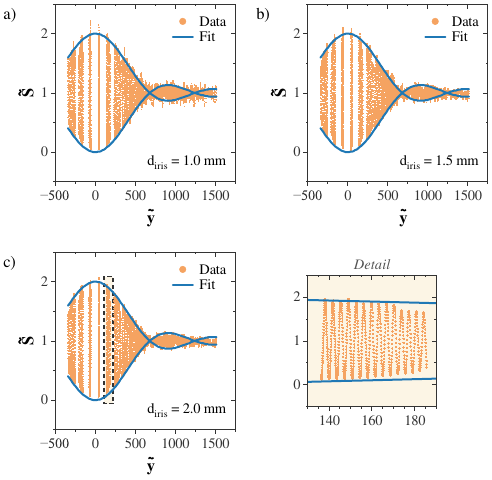}
    \caption{Results from classical thermal source, for different iris diameters of \SI{1.0}{\milli\meter} a), \SI{1.5}{\milli\meter} b). and \SI{2.0}{\milli\meter} c).
    The fit with the theoretical function gives a value of the source diameter of \SI{1.15}{\milli\meter}, \SI{0.97}{\milli\meter}, and \SI{0.97}{\milli\meter}, respectively, compatible with the nominal value of \SI{1.00}{\milli\meter}.
    In c) a detail of the measured reduced spectrum is shown on the right (detail from dashed box).}
    \label{fig:resclas}
\end{figure}

As far as the classical thermal source is concerned, Fig.~\ref{fig:specclas} shows an example of measurement with the spectrometer.
In this example, a vertical shift $\Delta y = \SI{40}{\micro\meter}$ is introduced with the crystal C$_2$, while a temporal delay of $\tau = \SI{93}{\femto\second}$ is introduced with the crystal C$_1$, giving rise to fringes.
In this case, after demodulating the source spectrum, we can retrieve a visibility of \SI{92.6}{\percent}.
We performed acquisitions for different values of $\Delta y$, then, we calculate the spectrum in reduced coordinates $\Delta \tilde{y} = \Delta y / \lambda$ and aggregate data.
Results are shown in Fig.~\ref{fig:resclas}.
Following Eq.~\ref{eq:specred}, the envelope of this trend is given by $\mu\left(\Delta \tilde{y}\right)$.
Assuming a circular shape of the source, we retrieve information about source size using Eq.~\ref{eq:mucirc}, in particular the radius $r$.
We performed measurements for different diameters of the interferometer iris (\SI{1.0}{\milli\meter} (a), \SI{1.5}{\milli\meter} (b), and \SI{2.0}{\milli\meter} (c)).
In the three cases, we obtained a value of the radius of \SI{1.15}{\milli\meter}, \SI{0.97}{\milli\meter}, and \SI{0.97}{\milli\meter}, respectively, compatible with the nominal value of \SI{1.00}{\milli\meter}.
The greater discrepancy of the first case is due to a lower signal on the spectrometer caused by the smaller iris aperture after the interferometer, so that zeroes of the Bessel function are noisy.
Notice that in the fit, we fixed the value of the focal length of the lens to \SI{500}{\milli\meter}.
These results confirm that the method is robust.

\begin{figure}
    \centering
    \includegraphics[width=82mm]{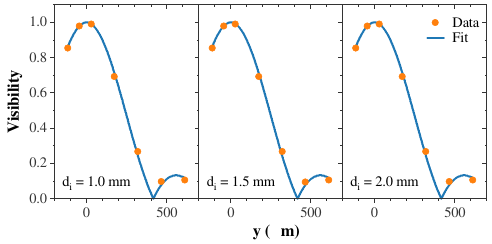}
    \caption{Results from interferograms of the classical thermal source.
    The fit with the theoretical function gives a value of the source diameter of \SI{0.996}{\milli\meter}, \SI{0.995}{\milli\meter}, and \SI{0.996}{\milli\meter}, respectively, compatible with the nominal value of \SI{1.00}{\milli\meter}.}
    \label{fig:resintf}
\end{figure}

A final measurement with classical light is acquiring interferograms instead of spectra.
In this part, the radiation after the interferometer is sent to a photon counter.
Acting on C$_1$, we measured the intensity on the detector with the photon counter as a function of $\tau$, obtaining fringes in the temporal domain.
Once spatial shift $\Delta y$ is fixed, their envelope is directly related to the temporal coherence of our source, while a constant factor is related to spatial coherence, according to Eq.~\ref{eq:cohetot}
Thus, by measuring the maximum visibility of each interferogram as a function of $\Delta y$, we could retrieve the spatial coherence function $\mu_S\left(\Delta y, \lambda\right)$ of Eq.~\ref{eq:cohetot}.
Results are reported in Fig.~\ref{fig:resintf}, for different iris diameters of the interferometer of \SI{1.0}{\milli\meter}, \SI{1.5}{\milli\meter}, and \SI{2.0}{\milli\meter}.
Assuming again a circular shape of the source, i.e., $\mu_S\left(\Delta y\right) = \mu_\mathrm{circ}(\Delta y,\lambda_0)$ (see Eq.~\ref{eq:mucirc}), being $\lambda_0$ the central wavelength of the radiation, we retrieve the source dimension of \SI{0.996}{\milli\meter}, \SI{0.995}{\milli\meter}, and \SI{0.996}{\milli\meter}, respectively.
These measurements are in good agreement with the nominal value of the source diameter of \SI{1}{\milli\meter} for each iris aperture.

\section{Conclusions}\label{sec:conc}
We have demonstrated a monolithic interferometer for spatial coherence measurements that operates effectively across both quantum and classical optical regimes. The device, based on two identical calcite crystals, successfully characterized spatial coherence properties of spontaneous parametric down-conversion (SPDC) photon pairs and classical thermal light from a halogen lamp. For the quantum source, measured spatial coherence lengths of $180\,\mathrm{\SIUnitSymbolMicro m}$ and $250\,\mathrm{\SIUnitSymbolMicro m}$ for pump waists of $840\,\mathrm{\SIUnitSymbolMicro m}$ and $560\,\mathrm{\SIUnitSymbolMicro m}$ respectively showed excellent agreement with theoretical predictions. Similarly, for the classical source, extracted diameters of $0.995$--$1.15\,\mathrm{mm}$ across different iris settings matched the nominal $1.00\,\mathrm{mm}$ pinhole size, confirming the method's accuracy.

The monolithic architecture provides inherent stability through rigid optical alignment, eliminating common experimental challenges associated with mechanical drift and thermal fluctuations. This design enables precise control of both temporal delays (via rotation angle $\alpha$ of crystal C$_1$) and spatial shearing (via angle $\gamma$ of crystal C$_2$) without requiring active stabilization systems. The interferometer's dual-mode detection capability, supporting both spectral analysis and coincidence counting, makes it possible to cross-validate measurements and ensures 
broad applicability to diverse light sources.

Our technique provides a unified platform for quantum-classical coherence characterization, bridging traditionally separate experimental domains. Additionally, the automated, computer-controlled implementation enables rapid measurements previously unattainable with conventional interferometers. Finally, the theoretical framework developed for SPDC correlations and thermal source coherence shows remarkable consistency with experimental data.

Our scheme may have applications in quantum imaging and sensing technologies where spatial coherence underpins performance limits. It may also offer practical utility in beam diagnostics for particle accelerators and astronomical instrumentation. Future research directions include extending the technique to pulsed sources, integrating the interferometer with integrated photonic circuits for miniaturization, and exploring spatiotemporal coherence coupling in complex light fields. The robust design principles demonstrated here may further enable adaptations for UV or X-ray coherence measurements using alternative birefringent materials.

\section*{References}
\bibliographystyle{iopart-num}
\bibliography{bibliography_NJP}

\end{document}